\documentclass[a4paper,UTF8,11pt,twoside]{cpc-hepnp}
\usepackage{multicol,multirow,graphicx,epsfig,subfigure,lastpage}
\usepackage{amssymb,bm,mathrsfs,bbm,amscd,amsfonts,booktabs}
\usepackage[tbtags]{amsmath}
\usepackage{CJK}
\usepackage{tabularx,latexsym,verbatim,color,braket}
\usepackage{dcolumn}
\usepackage{bm}
\usepackage[normalem]{ulem}
\newcommand{\Slash}[1]{\ooalign{\hfil/\hfil\crcr$#1$}}

\renewcommand\sout{\bgroup \color{red} \ULdepth=-.5ex \ULset}

\def\eq{\begin{eqnarray}}
\def\en{\end{eqnarray}}
\def\ds{d^*}
\def\bds{\bar{d}^*}
\def\dP{\bar{\rm P}}
\def\d12{D_{12}}
\def\cp{{\cal P}}
\def\tg{{\tilde g}}
\def\D{\Delta}
\begin{document}
\begin{CJK*}{GBK}{song}

\title{\bf  Possible dibaryon production at $\bar{\rm{P}}$anda with a Lagrangian approach
\thanks{We would like to thank Prof. Zongye Zhang for valuable discussions.
This work is supported by the National Natural Science Foundation of China under Grants
No.~11475192, the Sino-German CRC 110 "Symmetries and the Emergence of Structure in QCD" project by NSFC under the Grant
No.~12070131001, the Key Research Program of Frontier Sciences, CAS, under the Grant No.~Y7292610K1,
and the National Key Research and Development Program of China under Contracts No. 2020YFA0406300.
Supports from IHEP Innovation Fund under the grant No. Y4545190Y2 is also appreciated.}}

\author{Yubing Dong$^{1,2}$\email{Dongyb@ihep.ac.cn}%
\quad Pengnian Shen$^{1}$\email{Shenpn@ihep.ac.cn}} \maketitle
\address{%
$^1$ Institute of High Energy Physics, Chinese Academy of
Sciences, Beijing 100049, China\\
$^2$ School of Physical Sciences, University of Chinese
Academy of Sciences, Beijing 101408, China}

\begin{center}\date{\today}\end{center}

\begin{abstract}
In order to confirm the existence of the dibaryon state $d^*(2380)$ observed at WASA@COSY, we estimate
the production of the possible dibaryon and anti-dibaryon pair $\ds\bds$ at the energy region of the upcoming experiments at
$\dP$anda. Based on some qualitative properties of $\ds$ extracted from the analysez in the non-relativistic quark
model, the production cross section for this spin-3 particle pair are calculated with the help of an phenomenological
effective relativistic and covariant Lagrangian approach.\\
\end{abstract}

\maketitle

\section{Introduction}\label{Set:Set1}\vspace{0.2cm}\par\noindent\par

It is known that the history of the study of dibaryons, such as $H$ and $\ds$ particles, can be traced back to about 60 years ago (refer
to the review articles by Clement~\cite{Clement:2016vnl,Clement:2020mab}). Theoretically, the possible dibaryon states have been carefully
investigated with many approaches from the hadronic degrees of freedom (HDF) to the quark degrees of freedom (QDF). In 2009, the evidence of
$\ds$ was firstly reported by CELSIUS/WASA and WASA@COSY Collaborations~\cite{CELSIUS-WASA,
Adlarson:2011bh,Adlarson:2012fe,Adlarson:2014pxj}.  Observations of the existence of such a dibaryon were claimed in their series of
experiments. This is because that their observed peak cannot simply be understood by the role from either the intermediate Roper
excitation or from the t-channel intermediate $\Delta\Delta$ state, except by introducing a new intermediate resonance with its mass,
width, and quantum numbers being $2370\sim 2380~\rm{MeV}$, $70\sim 80~\rm{MeV}$, and $I(J^P)=0(3^+)$, respectively. Since the baryon
number of this resonance is 2, one believes it may just be the light-quark-only dibaryon $\ds(2380)$ that has been hunted for several decades.\\

According to the mass of $\ds$ and the relevant thresholds of the
two-baryon ($\D\D$), two-baryon plus one-meson ($NN\pi$), and
two-baryon plus two-meson ($NN\pi\pi$) channels near by, one could
believe the threshold (or cusp) effect should be much smaller in the
case of this resonance than that in the cases of the exotic XYZ
resonances~\cite{Chen:2016qju,Guo:2017jvc,Dong:2017gaw,Lebed:2016hpi}.
And due to its narrow width, one may think of this dibaryon as a
state with, at least, a hexaquark dominated structure. Up to now,
several theoretical proposals for its internal structure have been
investigated. Among them, two proposed structures have attracted
much attention. The first one comes from the study in QDF. The
calculation showed that $\ds$ has a compact structure and is a
candidate of an exotic hexaquark-dominated resonant
state~\cite{Yuan,Brodsky,Huang:2014kja,Huang:2015nja,Dong,Dong1,Dong2}.
(A similar assumption regards the $d^*$ state being a deeply bound
state of two $\Delta$s~\cite{Huang:2013nba,Huang:2018rpb}, however
its width is larger than the measured value.) The other one
considers it, in HDF, as a molecular-like hadronic state, which
originates from an assumption of a three-body resonance $\Delta
N\pi$ or a molecular-like state
$D_{12}\pi$~\cite{Gal:2013dca,Gal:2014zia,Platonova:2014rza,Platonova:2012am}.
Although the mass and the partial widths of the double pionic decays of such a hypothetic dibaryon resonance can be reasonably reproduced
by both proposed structures, the interpretations of it in two proposals are entirely different. Of course, there are
many other studies for understanding the structure of $\ds$ and  $pp\to \pi^+d$ reaction, for instance, the triple di-quark
model~\cite{Shi:2019dpo} and the triangle singularity  mechanism~\cite{Ikeno:2021frl}. Whether they can systematically
explain all existing experimental data still needs to be further tested. Therefore, it is necessary to look for other physical
observables in some sophisticated kinematics regions or some  processes other than $p$-$p$ (or $p$-d) collision, which might
explicitly provide remarkably different results for different structure models, especially the two $\ds$
structure models highlighted earlier. Actually, such theoretical analyses have been carried out on the electromagnetic form factors
of $\ds$~\cite{Dong:2017mio,Dong:2018emq} and on the possible evidence in the $\gamma+d$ processes~\cite{Dong:2019gpi}.\\

Up to now, the dibaryon $\ds$ has been observed by WASA@COSY Collaborations in the process of $pn\to d\pi\pi$ and the fusion
process of $pd\to ^3He+\pi\pi$. It seems that $\ds$ was also observed in another process, say $\gamma +d\to d \pi\pi$ at
ELPH~\cite{Ishikawa:2016yiq,Ishikawa:2018wkv,Ishikawa:2020jsh}. It should be stressed that the forthcoming experiments at
$\dP$anda (Pbar ANnihilation at DArmstadt) are expected to provide a confirmation of this dibaryon state if it does exist. This is
because that at $\dP$anda, the antiproton beam collides with the proton target and the momentum of $\bar{p}$ could be in the range
from $1$ GeV/c to 15 GeV/c.  It corresponds to a range of the total center-of-mass (CM) energy $\sqrt{s}$ of the proton-antiproton
system being from $\sim2.25~GeV$ to $\sim 5.5~GeV$~\cite{Fioravanti:2014vza,Hawranek:2007mp,Bettoni:2012zz}, which covers
$2M_{\ds}\sim 4.76~GeV$. Therefore, the future experiments based on the $p\bar{p}$ annihilation reaction can provide another way
to produce the dibaryon and anti-dibaryon pairs, say $\ds\bar{\ds}$, and can further give the information of this $\ds$ resonance. \\

In this work, a phenomenological effective Lagrangian approach (PELA) is employed to study the production of a spin-3 particle
$\ds$. It should be mentioned that this approach has been  successfully applied to many weakly bound state
problems~\cite{Dong:2017gaw} in the exotic meson sectors of $X(3872)$, $Z_b(10610)$, and
$Z_b(10650)$~\cite{Dong:2008gb,Dong:2009yp,Dong:2009uf,Dong:2013iqa} and the exotic baryon sector of
$\Lambda_c(2940)$~\cite{Dong:2009tg,Dong:2010xv}, and also  the  deuteron (S=1)~\cite{Dong:2008mt}. For the pion meson, which is different from
the above-mentioned loosely bound states, its properties can also be reasonably obtained by this approach~\cite{Faessler:2003yf}. Moreover,
this approach has been  applied to the study of the dibaryon candidate of $N \Omega$ (S=2)~\cite{Xiao:2020alj}, predicted by Ref.~\cite{Lis:2000}
and the HAL QCD collaboration~\cite{Iritani:2018sra}.  Therefore, as an extrapolation, PELA could be adopted as a reasonable tool to estimate the
cross section of $p\bar{p}\to \ds\bar{\ds}$ in the energy region of $\sqrt{s}\in [4.8,~5.50]~GeV$ at $\dP$anda.\\

This paper is organized as follows. In section 2, we show the description of the spin-3 dibaryon states by PELA. Then, a brief discussion for the
cross sections of the $p\bar{p}\to \Delta\bar{\Delta}$ and $p\bar{p}\to \Delta\bar{\Delta}\to \ds\bds$ processes is given in section 3. The
numerical results are displayed in section 4. Finally, section 5 is devoted for short summary and discussions.\\

\section{Description of the spin-3 particle $d^*(2380)$ in PELA}

By considering the interpretation of $\ds$ in the nonrelativistic quark model
in Refs.~\cite{Yuan,Huang:2014kja,Huang:2015nja,Dong,Dong1,Dong2}, here, we write the effective Lagrangian of $\ds~(3^+)$ and its
two constituents (for example two $\Delta$s) as
\eq
{\cal L}_{d^*\Delta\Delta}(x)&=&g_{_{\ds\Delta\Delta}}\int d^4y\Phi(y^2)\bar{\Delta}_{\alpha}(x+y/2)
\Gamma^{\alpha,(\mu_1\mu_2\mu_3),\beta}\Delta^C_{\beta}(x-y/2)d^*_{\mu_1\mu_2\mu_3}(x;\lambda)+\,H. C.\,
\en
where $\Delta_{\alpha}$ is the spin-3/2 $\Delta$ field, and $\Delta^C_{\alpha}$ stands for its charge-conjugated with
$\Delta^C_{\alpha}=C\bar{\Delta}^T_{\alpha}$ and  $C=i\gamma^2\gamma^0$. In the above equation, $d^*_{\mu_1\mu_2\mu_3}(x;\lambda)$
represents the spin-3 $\ds$ field with the polarization $\lambda$. It is a rank-3 field. The coupling of the two $\Delta$s to $\ds$
relates to the two spin-3/2 particles and a spin-3 particle. The three-particle vertex reads~\cite{Scadron:1968zz}
\eq
\Gamma^{\alpha,(\mu_1\mu_2\mu_3),\beta}&=&\frac16\Big [\gamma^{\mu_1} \Big
(g^{\mu_2\alpha}g^{\mu_3\beta}+g^{\mu_2\beta}g^{\mu_3\alpha}\Big )+\gamma^{\mu_2}
\Big (g^{\mu_3\alpha}g^{\mu_1\beta}+g^{\mu_1\beta}g^{\mu_3\alpha}\Big )\\ \nonumber
&&~~+\gamma^{\mu_3} \Big (g^{\mu_1\alpha}g^{\mu_2\beta}+g^{\mu_1\beta}g^{\mu_2\alpha}\Big )\Big ].
\en
The introduced correlation function $\Phi(y^2)$ in eq. (1) describes the distribution of the two constituents in the system and makes the
integral of Feynman diagrams finite in ultraviolet. This function is related to its Fourier transform in momentum space, $\tilde\Phi(-p^2)$,
by $\Phi(y^2)=\int \frac{d^4 p}{(2\pi)^4}e^{-ipy}\tilde\Phi(-p^2)\,$ where $p$ stands for the relative Jacobi momentum between the two
constituents of $\ds$. For simplicity, $\tilde\Phi$ is phenomenologically chosen in a Gaussian-like form as
\eq
\tilde\Phi(-p^2) = \exp(p^2/\Lambda^2),
\en
where $\Lambda$ is a model parameter, relating to the scale of the distribution of the constituents inside $d^*$, and has dimension of mass.
All calculations for the loop integral, hereafter, are performed in Euclidean space after the Wick transformation, and all the external momenta
go like $p^\mu=(p^0,\vec{p\,}) \to p^\mu_E=(p^4,\vec{p\,})$ (where the subscript "$E$" stands for the momentum in Euclidean space) with
$p^4 = -ip^0$. In Euclidean space the Gaussian correlation function ensures that all loop integrals are ultraviolet finite
(details can be found in Ref.~\cite{Dong:2017gaw}).\\

Then, one can determine the coupling of $\ds$ to its constituents by using the Weinberg-Salam compositeness
condition~\cite{Salam:1962ap,Weinberg:1962hj,Hayashi:1967hk,Efimov:1993ei}.  This condition means that the probability of finding the dressed
bound state as a bare (structureless) state is equal to zero.  In  the case of $\ds$, our previous calculation in
QDF~\cite{Yuan,Huang:2014kja, Huang:2015nja,Dong,Dong1,Dong2} shows  that $\ds$ contains a $|\Delta\Delta>$ and also a $|CC>$ components, which are
orthogonal to each other.  As a rough estimation, a simplest chain approximation is used.  Then this condition can be written as
\eq
Z_{\ds} = 1 - \frac{\partial\Sigma^{(1)}_{(\D\D)}(\cp^2)} {\partial\cp^2}\Big
|_{\cp^2=M^2_{\ds}}- \frac{\partial\Sigma^{(1)}_{(CC)}(\cp^2)} {\partial\cp^2}\Big
|_{\cp^2=M^2_{\ds}}=Z_{\ds, (\D\D)}+Z_{\ds, (CC)}=0\,,
\label{eq:compositeness-con-1}
\en
where $\cp$ is the momentum of $\ds(2380)$,
$\Sigma^{(1)}_{(\D\D)~\text{or}~(CC)}(M^2_{\ds})$ is
the non-vanishing part of structural integral of the mass operator of $\ds$ with spin-parity $3^+$ (the detailed derivation can be found in
Refs.~\cite{Hayashi:1967bjx,Efimov:1987sa}). Here we assume these $Z_{\ds, (\D\D)}$ and $Z_{\ds, (CC)}$ are independent. Since the
probabilities of the $\D\D$ and $CC$ components
are about $P_{\D\D}\sim 1/3$ and $P_{CC}\sim 2/3$, respectively in quark model calculation, therefore,
\begin{align}
Z_{\ds, (\D\D)}=\frac13-\frac{\partial\Sigma^{(1)}_{(\D\D)}(\cp^2)} {\partial\cp^2}\Big
|_{\cp^2=M^2_{\ds}}=0,\tag{4a}\\ \nonumber
Z_{\ds, (CC)}=\frac23-\frac{\partial\Sigma^{(1)}_{(CC)}(\cp^2)} {\partial\cp^2}\Big
|_{\cp^2=M^2_{\ds}}=0, \tag{4b}
\end{align}
and the coupling $g_{_{\ds\Delta\Delta}}$ can be extracted from the compositeness condition of (4a). The mass operator of the $\ds$ dressed by
the $\Delta\Delta$ channel is given in Fig. 1. It should be
stressed that the coupling constant determined from the compositeness condition of eq.~(4a) contains the renormalization
effect since the chain approximation is considered (also refer to Refs.~\cite{Hayashi:1967bjx,Efimov:1987sa}).\\

\begin{figure}
\centering
\includegraphics [width=10cm, height=5cm]{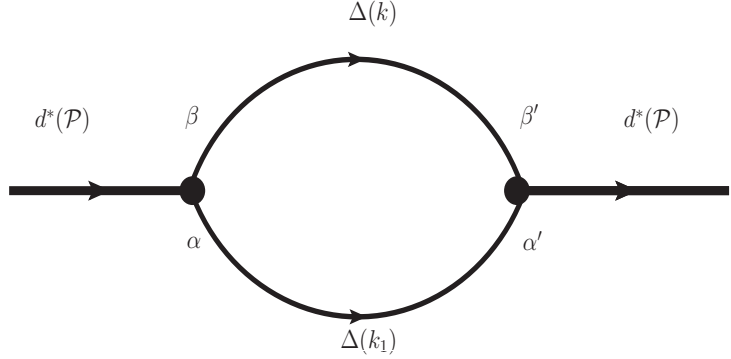}
\caption{The Mass operator of the $d^{*}(2380)\to\Delta\Delta$. }
\end{figure}

The explicit expression of the full mass operator can be written as
\eq
&&\Sigma_{\D}^{(\mu_i'), (\mu_j)}({\cal P})=\big| g_{_{\ds\D\D}}(\Lambda)\big|^2\int \frac{d^4k}{(2\pi)^4i}
\exp\Big (-\frac{2(k-\frac{{\cal P}}{2})_E^2}{\Lambda^2}\big )\times Tr\Bigg\{\Gamma_{\alpha'~~~\beta'}^{~(\mu_i')}\\ \nonumber
&&~~~~~~~~\times\frac{\Slash{k}+M_{\D}}{k^2-M_{\D}^2} \big (-g^{\beta\beta'}+\frac{\gamma^{\beta}\gamma^{\beta'}}{3}+
\frac{2k^{\beta}k^{\beta'}}{3M_{\D}^2} +\frac{\gamma_{\beta}k_{\beta'}-\gamma^{\beta'}k^{\beta}}{3M_{\D}}
\big )\times\Gamma_{\beta~~~\alpha}^{~(\mu_j)}\\ \nonumber
&&~~~~~~~~\times\frac{\Slash{k}_1-M_{\D}}{k_1^2-M_{\D}^2} \big (-g^{\alpha'\alpha}+\frac{\gamma^{\alpha'}\gamma^{\alpha}}{3}+
\frac{2k^{\alpha'}_{1}k^{\alpha}_{1}}{3M_{\D}^2} -\frac{\gamma^{\alpha'}k^{\alpha}_{1}-\gamma^{\alpha}k^{\alpha}_{1}}{3M_{\D}}
\big )\Bigg \}\Bigg |_{k_1={\cal P}-k},
\en
with $\mu_i'$ and $\mu_j$ being the abbreviations of $(\mu_1',\mu_2',\mu_3')$ and $(\mu_1,\mu_2,\mu_3)$, respectively. In general, according
to its Lorentz structure, the mass operator $\Sigma_{(c)}^{(\mu_i'),(\mu_j)}({\cal P})$ takes the form of
\eq
\Sigma_{(c)}^{(\mu_i'),(\mu_j)}({\cal P})=\sum_{l=1}^8
L^{(\mu_i'), (\mu_j)}_{(l),(c)}\Sigma^{(l)}_{(c)}({\cal P}^2),\,
\en
with $\Sigma^{(l)}_{(c)}({\cal P}^2)$ being the structural integrals appeared in the expression of the full mass operator, and the Lorentz
structures being
\eq
L^{(\mu_i'), (\mu_j)}_{(1)}&=&\frac{1}{6}\big [g^{\mu_1'\mu_1} \big
(g^{\mu_2'\mu_2}g^{\mu_3'\mu_3}+g^{\mu_2'\mu_3}g^{\mu_3'\mu_2}\big )+g^{\mu_1'\mu_2} \big
(g^{\mu_2'\mu_1}g^{\mu_3'\mu_3}+g^{\mu_2'\mu_3}g^{\mu_3'\mu_1}\big )\\ \nonumber
&&~~~+g^{\mu_1'\mu_3}\big
(g^{\mu_2'\mu_2}g^{\mu_3'\mu_1}+g^{\mu_2'\mu_1}g^{\mu_3'\mu_2}\big )\big ]\,,\\ \nonumber
&=&\frac16 \big [g^{\alpha_1\beta_1}(g^{\alpha_2\beta_2}g^{\alpha_3\beta_3}+g^{\alpha_2\beta_3}g^{\alpha_3\beta_2})+......\big ]
\en
with $\alpha_i\in (\mu_1',\mu_2',\mu_3')$ and $\beta_j\in (\mu_1,\mu_2,\mu_3)$,
\eq
&&L^{(\mu_i'), (\mu_j)}_{(2)} =\frac19\big
[g^{\alpha_1\alpha_2}\big
(g^{\alpha_3\beta_1}g^{\beta_2\beta_3}+g^{\alpha_3\beta_2}g^{\beta_3\beta_1}
+g^{\alpha_3\beta_3}g^{\beta_1\beta_2}\big )+......\big ]\,, \en \eq
L^{(\mu_i'), (\mu_j)}_{(3)}=\frac{1}{18}\big
[\cp^{\alpha_1}\cp^{\beta_1} \big
(g^{\alpha_2\beta_2}g^{\alpha_3\beta_3}+g^{\alpha_2\beta_3}g^{\alpha_3\beta_2}\big
)+...... \big ]\,, \en \eq L^{(\mu_i'), (\mu_j)}_{(4)} =\frac19\big
[\cp^{\alpha_1}\cp^{\beta_1}\big
(g^{\alpha_2\alpha_3}g^{\beta_2\beta_3}\big )+...... \big ]\,, \en
\eq L^{(\mu_i'), (\mu_j)}_{(5)} &=&\frac{1}{18}\big
[\cp^{\alpha_1}\cp^{\alpha_2}\big
(g^{\alpha_3\beta_1}g^{\beta_2\beta_3}+
g^{\alpha_3\beta_2}g^{\beta_1\beta_3}+g^{\alpha_3\beta_3}g^{\beta_1\beta_2}\big
)+......\\ \nonumber &&~~+\cp^{\beta_1}\cp^{\beta_2}\big
(g^{\alpha_1\beta_3}g^{\alpha_2\alpha_3}+
g^{\alpha_2\beta_3}g^{\alpha_2\alpha_3}+g^{\alpha_3\beta_3}g^{\alpha_1\alpha_2}\big
)+......\big ]\,, \en \eq L^{(\mu_i'), (\mu_j)}_{(6)}
=\frac{1}{9}\big
[\cp^{\alpha_1}\cp^{\alpha_2}\cp^{\beta_1}\cp^{\beta_2}\big
(g^{\alpha_3\beta_3}\big )+...... \big ]\,, \en \eq L^{(\mu_i'),
(\mu_j)}_{(7)} =\frac{1}{6}\big
[\cp^{\alpha_1}\cp^{\alpha_2}\cp^{\alpha_3}\cp^{\beta_1}\big
(g^{\beta_2\beta_3}\big )+
\cp^{\beta_1}\cp^{\beta_2}\cp^{\beta_3}\cp^{\alpha_1}\big
(g^{\alpha_2\alpha_3}\big )+......\big ]\,, \en \eq L^{(\mu_i'),
(\mu_j)}_{(8)}=\cp^{\mu_1'}\cp^{\mu_2'}\cp^{\mu_3'}\cp^{\mu_1}\cp^{\mu_2}\cp^{\mu_3}.
\en
Clearly, due to the property of the polarization vector of the spin-3 particle, like $\epsilon_{\mu_1\mu_2\mu_3}(\cp,\lambda)$ shown in
Ref.~\cite{Dong:2017mio}, only the first term on the right-hand side of eq. (6) gives the contribution while the other terms
do not. We introduce Lorentz projector
\eq
&&T_{\perp}^{(\mu_i'),(\mu_j)}=\frac{1}{42} \big
[\tg^{\mu_1'\mu_1}\big
(\tg^{\mu_2'\mu_2}\tg^{\mu_3'\mu_3}+\tg^{\mu_2'\mu_3}\tg^{\mu_3'\mu_2}\big
)\\ \nonumber &&~~~~~~~~~~~~~~~~~~+\tg^{\mu_1'\mu_2}\big
(\tg^{\mu_2'\mu_1}\tg^{\mu_3'\mu_3}+\tg^{\mu_2'\mu_3}\tg^{\mu_3'\mu_1}\big
) +\tg^{\mu_1'\mu_3}\big
(\tg^{\mu_2'\mu_2}\tg^{\mu_3'\mu_1}+\tg^{\mu_2'\mu_1}\tg^{\mu_3'\mu_2}\big
)\big ]\\ \nonumber &&~~~~~~~~~~~~~~~~~~-\frac{1}{105}\big
[\tg^{\mu_1'\mu_2'}\big (\tg^{\mu_3'\mu_1}\tg^{\mu_2\mu_3}
+\tg^{\mu_3'\mu_2}\tg^{\mu_1\mu_3}+\tg^{\mu_3'\mu_3}\tg^{\mu_1\mu_2}\big
)\\ \nonumber &&~~~~~~~~~~~~~~~~~~+\tg^{\mu_1'\mu_3'}\big
(\tg^{\mu_2'\mu_1}\tg^{\mu_2\mu_3}
+\tg^{\mu_2'\mu_2}\tg^{\mu_1\mu_3}+\tg^{\mu_2'\mu_3}\tg^{\mu_1\mu_2}\big
)\\ \nonumber &&~~~~~~~~~~~~~~~~~~+\tg^{\mu_2'\mu_3'}\big
(\tg^{\mu_1'\mu_1}\tg^{\mu_2\mu_3}
+\tg^{\mu_1'\mu_2}\tg^{\mu_1\mu_3}+\tg^{\mu_1'\mu_3}\tg^{\mu_1\mu_2}\big
)\big ]\,, \en with \eq
\tg^{\mu\nu}=g_{\perp}^{\mu\nu}=-g^{\mu\nu}+\frac{{\cal
P}^{\mu}{\cal P}^{\nu}}{M^2_{\ds}}.
\en
It satisfies following relations
\eq
{\cal P}_iT_{\perp}^{(\mu_i'),(\mu_j)}=0,
~~~~\mu_i'\in (\mu_1',\mu_2',\mu_3')~\rm{or}~\mu_j\in
(\mu_1,\mu_2,\mu_3); \en \eq L_{(\mu_i'),
(\mu_i)}^{(1)}T_{\perp}^{(\mu_i'),(\mu_j)}=1 \en and \eq
L_{(\mu_i'),
(\mu_j)}^{(i)}T_{\perp}^{(\mu_i'),(\mu_j)}=0,~~~~~(i=2,3,...,8).
\en
Thus, when the full mass operator $\Sigma^{(\mu_i'),(\mu_j)}({\cal P})$ acts with the Lorentz projector $T_{\perp}^{(\mu_i'),(\mu_j)}$, the
product gives the scalar function $\Sigma^{(1)}({\cal P}^2)$ in eq.(4), and  it will contribute to the compositeness
condition. Finally the coupling constant $|g|^2_{_{\ds\Delta\Delta}}$ can be determined from eq. (4a).\\

It should be stressed that here we have adopted the Gaussian-type correlation function of eq.(3), $\tilde\Phi(-p^2) = \exp(p^2/\Lambda^2)$,
the model-dependent parameter $\Lambda$ relates to the size of the system in the non-relativistic approximation, at least in physical meaning.
Thus, one may roughly connect $b$, representing the size of the $\ds$ in the non-relativistic wave function, to the parameter $\Lambda$  by
$\frac{b^2}{2}\sim \frac{1}{\Lambda^2}$. According to the quark model calculation in Refs.~\cite{Huang:2015nja}, $b\sim 0.8fm$, and we roughly
choose the parameter $\Lambda\sim 0.34~GeV$. \\

\section{Cross sections for $p\bar{p}\to \Delta\bar{\Delta}$ and $p\bar{p}\to\Delta\bar{\Delta}\to \ds\bar{\ds}$}

\subsection{Cross section for $p\bar{p}\to \Delta\bar{\Delta}$}

There are only a few experiments of the $p\bar{p}\to \Delta(1232)\bar{\Delta}(1232)$ process in the literature
~\cite{VanApeldoorn:1978jd,Ward:1978npb,Johnson:1980nw,VanApeldoorn:1982iw,Saleem:1983ky}. Refs.~\cite{VanApeldoorn:1982iw,Saleem:1983ky}
studied $p\bar{p}\to \Delta\bar{\Delta}$ at $7.23~GeV$ and $12~GeV$. The samples were obtained from the large exposures of the 2m
hydrogen bubble-chamber (HBC) experiment to the U5 antiproton beam at CERN.  The account of $p\bar{p}\pi^+\pi^-$ was thought to come
dominantly from the $\D^{++}\overline{\D^{++}}$ channel. It was believed that the process can be described by the t-channel pion or
reggeized pion exchange. A good description of the mass and t-distributions for the reaction at $3.6~GeV$ and $5.7~GeV$ was given by
the one-pion exchange model~\cite{Wolf:1969iw}. Moreover the cross section of the process, in terms of the Mandelstam variable of $s$,
is parameterized as $\sigma (s)=As^{-n}$ with $A=(67\pm 20)~mb$ and $n=1.5\pm 0.1$, respectively~\cite{VanApeldoorn:1982iw}.\\

This $p\bar{p}\to \Delta\bar{\Delta}$ process can also be estimated theoretically by using an effective Lagrangian~\cite{Cao:2010ji}
\eq
{\cal L}^{(t_{z}^{\Delta}t_z^N)}_{\pi N\Delta}=g_{_{\pi N\Delta}}F(p_t)\bar{\D}_{\mu}^{(t_z^{\D})}\vec{\cal I}_{t_z^{\D}t_z^{N}}\cdot\partial^{\mu}
\vec{\pi}^{(t_z^{\pi})}N^{(t_z^{N})}~+~h.c.,
\en
where $g_{_{\pi N\Delta}}$ and $F(p_t)$ are the effective coupling constant and phenomenological form factor, respectively, the
latter function is chosen to be
\eq
F(p_t)=\Big(\frac{\Lambda_M^{*2}-m_{\pi}^2}{\Lambda_M^{*2}-p_t^2}\Big )^n\exp(\alpha p_t^2),
\en
with the parameters $\Lambda_M^{*}\sim 1~GeV$ and $n=1$. In eq. (20),
$\vec{\cal I}_{t_z^{\D}t_{z}^{N}}=C_{1t_{\pi},1/2t_z^N}^{3/2t_z^{\D}}\hat{e}^*_{t_{\pi}}$ is the isospin transition operator.
Then, the cross section is
\eq
\sigma=
\int\frac{(2\pi)^4\delta^4(p_1+p_2-p_3-p_4)}{4\sqrt{(p_1\cdot
p_2)-m_1^2m_2^2}} \times \sum_{Pol.}\Big |\overline{\cal M}_{if}\Big
|^2\frac{d^3p_3}{(2\pi)^32E_{p_3}}\frac{d^3p_4}{(2\pi)^32E_{p_4}},
\en
where $p_{1,2}$ (or $p_{3,4}$) are the momenta of the incoming (or outgoing) particles, $\Big |\overline{\cal M}_{if}\Big |^2$
stands for averaging over the polarizations of the initial states and summing over the polarizations of final states. We can write
the matrix element $\overline{\cal M}_{if}$, representing the contribution of the tree-diagram to $p\bar{p}\to \Delta^{++}\overline{\D^{++}}$,
via $\pi$ exchange with the Lagrangian of eq. (20), as
\eq
{\cal M}_{if}^{p\bar{p}\to \D^{++}\overline{\D^{++}}}=g^2_{_{\pi N\Delta}}F^2(p_t)\Big [\bar{U}^{\D}_{\alpha}p_t^{\alpha}u(p_1)\Big ]
\frac{1}{p_t^2-m_{\pi}^2}\Big [\bar{v}(p_2)p_t^{\beta}V^{\D}_{\beta}(p_4)\Big ].
\en
\par\noindent
{\hskip 0.4cm} The resultant cross section is shown and compared with the parameterized empirical cross section in Fig. 2.
It should be mentioned that in Ref.~\cite{Riska:2000gd} $F_{_{\pi N\Delta}}=f_{_{\pi N\Delta}}/m_{\pi}$, and in order to fit the decay
width of $\D\to \pi N$, where the initial momentum of $\D$ is set to be zero, the value of $f_{_{\pi N\D}}$ is taken as $2.2\pm 0.04$.
Thus, their $F_{_{\pi N\D}}\sim (15.7\pm 0.285)~ GeV^{-1}$. In our present numerical calculation, to fit the parameterized cross
section, we introduce an addition trajectory function $\exp(0.2t)$ $(p_t^2=t~<~0)$ and take $F_{_{\pi N\Delta}}\sim 10.75~GeV^{-1}$. Here,
we find that the $10\%$ variation in $F_{_{\pi N\Delta}}$ may cause about $50\%$ change in the total cross section since the cross section
is proportional to $F^4_{_{\pi N\D}}$. In addition, the change of the estimated $\sqrt{s}$-dependent cross section with respect to the
variations of the parameters $\Lambda^*_M$ and $\alpha$ are shown in this figure as well. Those curves tell that the cross section
with smaller $\sqrt{s}$ becomes larger when $\Lambda^*_M$ deceases or $\alpha$ increases. The combined effect of $\Lambda^*_M$ and $\alpha$,
namely the effect of the phenomenological form factor, on the cross section is more pronounced in the small $\sqrt{s}$ region. Therefore,
the current lagrangian is flexible enough to fit the experimental data. It should be mentioned that in this calculation, we only
consider the one-pion exchange, insert a phenomenological form factor, and take the coupling of $F_{_{\pi N\Delta}}$ as a free parameter.
It seems that our tree diagram result is reasonable to reproduce the total cross section of $p\bar{p}\to \D^{++}\bar{\D^{++}}$, although we
do not consider the contributions from other meson exchange, for instance the $\rho$ meson. In conclusion, the effective Lagrangian
${\cal L}^{(t_{z}^{\Delta}t_z^N)}_{_{\pi N\Delta}}$ mentioned above is appropriate for describing the cross section of the
$p\bar{p}\to \Delta\bar{\Delta}$ process, so it should also be acceptable and reasonable to be further used in the investigation of the
$d^*\bar{d^*}$ generation in the $p\bar{p}\to\Delta\bar{\Delta}\to \ds\bar{\ds}$ process.\\

\vspace{0.5cm}
\begin{figure}[ht]
\centering
\includegraphics [width=8cm, height=6cm]{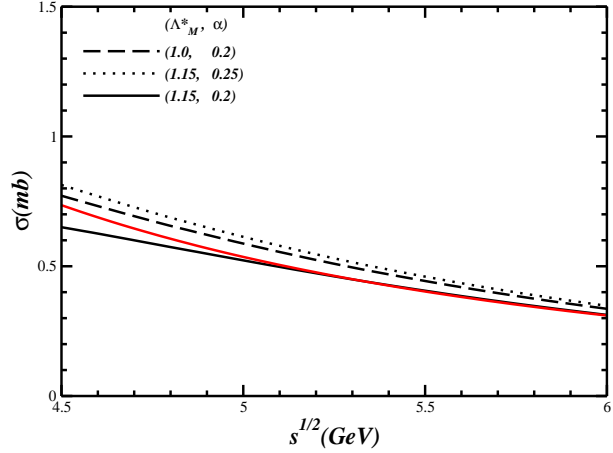}\vspace{0.5cm}
\caption{The estimated cross sections for $p\bar{p} \to\Delta^{++}\overline{\Delta^{++}}$ compared to
that with a parameterized form of $\sigma (\text{mb})=67s^{-1.5}$ (red curve). The black solid, dashed, and dotted curves represent the
calculated results with the parameters of $(\lambda^*_{M}(GeV),\alpha(GeV^2))$ being (1.15, 0.2), (1.0, 0.2), and (1.15, 0.25), respectively.}
\end{figure}

\subsection{Cross section for $p\bar{p}\to\Delta\bar{\Delta}\to \ds\bar{\ds}$}

\begin{figure}
\centering
\includegraphics [width=9cm, height=6cm]{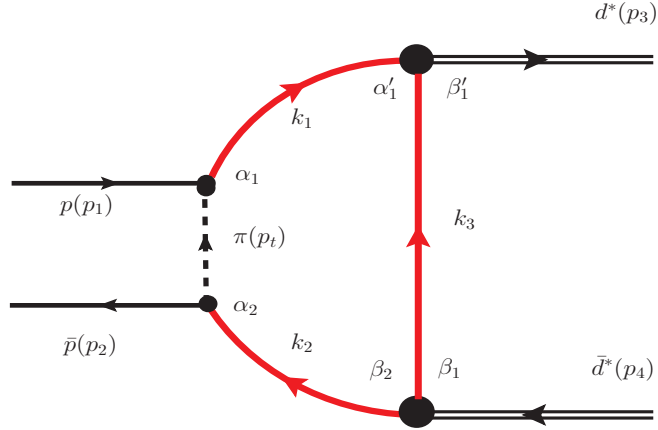}\vspace{0.5cm}
\caption{The Feynman diagram for the $p\bar{p}\to
d^*(2380)+\bar{d}^*(2380)$ process, where the red bold line and
double line stand for the internal $\D$ (or $\D^C$) field and the outgoing
$\ds$, respectively. }
\end{figure}

The Feynman diagram of the $p\bar{p}\to \ds\bar{\ds}$ process via $\Delta\bar{\D}$ intermediate is shown in Fig. 3. In this diagram,
$\ds\bar{\ds}$ pair is generated from the $p\bar{p}\to\Delta\bar{\Delta}$ annihilation reaction.
It should be noted that in the loop, in the higher order approximation, when $p\bar{p}$ annihilation generates a
$\Delta\bar{\Delta}$ pair, it can also create a corresponding $C\bar{C}$ pair, therefore, when $\Delta$ interacts with $\Delta$
(or $\bar{\Delta}$ interacts with $\bar{\Delta}$), a corresponding hidden-color component $CC$ (or $\bar{C}\bar{C}$) would exist.
According to the conclusion in our previous quark model calculations, about 1/3 of $\Delta\Delta$ ($\bar{\Delta}\bar{\Delta}$) and 2/3
of $CC$ ($\bar{C}\bar{C}$) can form a $d^*$ ($\bar{d^*}$),as
\begin{eqnarray*}
|\ds>\sim \sqrt{\frac{1}{3}}|\D\D>+\sqrt{\frac{2}{3}}|CC>,
\end{eqnarray*}
with the spin and isospin quantum numbers of the colored cluster $C$ being 3/2 and 1/2. Thus, to estimate events of $\ds$
($\bar{\ds}$) creation, we can only use 1/3 of $\Delta\Delta$  ($\bar{\Delta}\bar{\Delta}$) component, because it corresponds to
one $d^*$ ($\bar{d^*}$). It should be further stressed that the process in this diagram can occur only when the Mandelstam variable
satisfies $\sqrt{s}>2M_{\ds}\sim 4.8~GeV$. It is clear that the threshold of this production channel is lower than the upper limit
of the CM energy of the $\dP$anda device. \\

To calculate the matrix element of Fig. 3, we have to use the vertices of ${\cal L}_{\pi N\Delta}$ in eq. (20) and ${\cal L}_{\ds\Delta\Delta}$
in eq. (1). The matrix element of ${\cal M}_{if}$ for the process of $p\bar{p}\to\ds\bar{\ds}$ reads
\eq
{\cal M}_{if}^{(p\bar{p}\to\ds\bar{\ds})}&=&\bar{v}_N(p_2)\Pi_{(\nu_i),(\mu_j)}u_N(p_1)\big (d^*(p_3)
\big )^{(\mu_j)}(\lambda)\big (\bar{d}^*(p_4)\big )^{(\nu_i)}(\bar{\lambda}),
\en
with
\eq \Pi_{(\nu_i),(\mu_j)}
&=&\int\frac{d^4p_t}{(2\pi)^4i}p_t^{\alpha_2}S^C_{3/2,(\alpha_2\beta_2)}(k_2)\Gamma_{\beta_2,(\nu_i),\beta_1}
S^C_{3/2,(\beta_1\beta_1')}(k_3)\\ \nonumber
&&\times \Gamma_{\beta'_1,(\mu_j),\alpha'_1}S_{3/2,(\alpha_1'\alpha_1)}(k_1)p_{t,\alpha_1}\frac{F^2(p_t)}{p_t^2-m_{\pi}^2}\\ \nonumber
&&~\times \exp\Big [-\Big (\frac{(k_1-k_3)^2_E}{4\Lambda^2}+\frac{(k_2-k_3)_E^2}{4\Lambda^2}\Big )\Big ]\times C_{Iso},
\en
where the exponential factors in the last arrow on the right side of eq. (25) come from the consideration of
the phenomenological bound state problem of $\ds$ discussed explicitly in section 2, and the subscripts "$E$" and "$M$" denote
"Euclidean" and "Minkowski", respectively. The propagators of a spin-3/2 particle $\Delta$ and its charge conjugate are
\eq
S_{3/2,\mu\nu}(p,M_{\D}) &=&(p\!\!\!\slash-m)^{-1}  \times \Big(-g_{\mu\nu}+\frac{\gamma_\mu\gamma_\nu}{3}
+\frac{2p_\mu p_\nu}{3M_{\D}^2} +\frac{\gamma_\mu p_\nu-\gamma_\nu p_\mu}{3M_{\D}}\Big ) \,,\\ \nonumber
S^C_{3/2,\nu\mu}(p,M_{\D}) &=& CS^{T}_{3/2,\mu\nu}(p,M_{\D}) C \,,
\en
with the charge conjugate operator being $C=i\gamma^2\gamma^0$. Moreover, the constant $C_{Iso.}=\frac{7}{18}$ represents the isospin factor
since the intermediate state can be either $\D^{++}\overline{\D^{++}}$, or $\D^+\overline{\D^+}$, or $\D^0\overline{\D^0}$
(here we only consider the pion-exchange in the $p\bar{p}\to \D\bar{\D}$ process). Then, the cross section of such a process is formally
expressed by eq. (22), where the matrix element is replaced by ${\cal M}_{if}^{(p\bar{p}\to\ds\bar{\ds})}$ given in eq. (24).
Noticed that the square of the matrix element is proportional to $g^4_{_{\ds\D\D}}$ and $g^4_{_{\pi N\D}}$, respectively.
Here, since the $\ds$ is a spin-3 particle, its field can be described by a traceless rank-3 polarization
vector like $\epsilon_{\mu_1\mu_2\mu_3}(\cp,\lambda)$. This polarization vector has the properties of $\epsilon_{\alpha\alpha\beta}=0$,
$\epsilon_{\alpha\beta\gamma}=\epsilon_{\beta\alpha\gamma}$, and  $\cp^{\alpha}\epsilon_{\alpha\beta\gamma}=0$. Therefore, in the
summation calculation, we have
\eq
\sum_{pol.}\epsilon_{\mu\nu\sigma}\epsilon^*_{\alpha\beta\gamma}
&=&\frac16\Big [{\tilde g}_{\mu\alpha}\Big ({\tilde
g}_{\nu\beta}{\tilde g}_{\sigma\gamma} +{\tilde
g}_{\nu\gamma}{\tilde g}_{\sigma\beta}\Big ) +{\tilde
g}_{\mu\beta}\Big ({\tilde g}_{\nu\alpha}{\tilde g}_{\sigma\gamma}
+{\tilde g}_{\nu\gamma}{\tilde g}_{\sigma\alpha}\Big ) +{\tilde
g}_{\mu\gamma}\Big ({\tilde g}_{\nu\alpha}{\tilde g}_{\sigma\beta}
+{\tilde g}_{\nu\beta}{\tilde g}_{\sigma\alpha}\Big ) \Big ]\\
\nonumber &&~~~-\frac{1}{15}\Big [{\tilde g}_{\mu\nu}\Big ({\tilde
g}_{\sigma\alpha}{\tilde g}_{\beta\gamma} +{\tilde
g}_{\sigma\beta}{\tilde g}_{\alpha\gamma} +{\tilde
g}_{\sigma\gamma}{\tilde g}_{\alpha\beta}\Big ) +{\tilde
g}_{\mu\sigma}\Big ({\tilde g}_{\nu\alpha}{\tilde g}_{\beta\gamma}
+{\tilde g}_{\nu\beta}{\tilde g}_{\alpha\gamma} +{\tilde
g}_{\nu\gamma}{\tilde g}_{\alpha\beta}\Big )\\ \nonumber
&&~~~~~+{\tilde g}_{\nu\sigma}\Big ({\tilde g}_{\mu\alpha}{\tilde
g}_{\beta\gamma} +{\tilde g}_{\mu\beta}{\tilde g}_{\alpha\gamma}
+{\tilde g}_{\mu\gamma}{\tilde g}_{\alpha\beta}\Big ) \Big ],
\en
with ${\tilde g}_{\mu\nu}$ showed in eq. (16). \\

\section{Numerical results and discussions}

\subsection{Discussion of $g_{_{\ds\Delta\Delta}}$}

In this work, we employ our phenomenological effective Lagrangian approach to describe the spin-3 resonance $\ds$.
Consequently, the Feynman diagram of Fig. 3 can be calculated covariantly and relativistically. In the calculation,
there is only one unique model parameter $\Lambda$. We fix this parameter according to the qualitative conclusions obtained from the
dynamical calculation in the non-relativistic constituent quark model~\cite{Huang:2015nja,Dong}: (1) $\ds$ contains two components
$|\Delta\Delta>$ and $|CC>$ with probabilities $1/3$ and $3/2$, respectively; (2) $\ds$ is a compact system with a size about
$b\sim 0.8~fm$; and (3) In the quark model approach, the strong decay widths of $\ds$, in the leading order approximation, are
dominantly contributed by the $\Delta\Delta$ component. Thus, $\Lambda^2 \sim \frac{2}{b^2}$, which gives $\Lambda \sim 0.34~GeV$ when
$b\sim 0.8~fm$. Further taking $P_{\Delta\Delta} \sim 1/3$, we can calculate the coupling constant of $\ds$ to $\Delta\Delta$ by
using the formulas shown in section 2. The result shows $g_{_{\ds\Delta\Delta}}\sim 3.35$. We present the change of the dimensionless
coupling constant $g_{_{\ds}}=g_{_{\ds\Delta\Delta}}\big /\sqrt{P_{\Delta\Delta}}$  with respect to the variation of the model-parameter
$\Lambda$ in the region of $[0.25,~0.45]~GeV$ in Fig. 4. \\

\vspace{0.5cm}
\begin{figure}[ht]
\centering
\includegraphics [width=8cm, height=6cm]{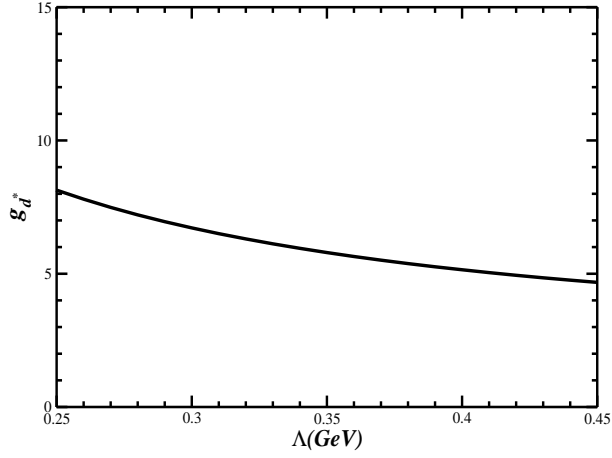}\vspace{0.5cm}
\caption{$g_{_{\ds}}=g_{_{\ds\Delta\Delta}}\big /\sqrt{P_{\Delta\Delta}}$ in PELA versus $\Lambda \in [0.25,~0.45]~GeV$. }
\end{figure}

The curve in Fig. 4 shows that the dimensionless coupling $g_{\ds}$ relates to the model parameter $\Lambda$ and to the integral of the
mass operator structure. When $\Lambda$ increases, the integral of the loop structure increases, and consequently, the obtained $g_{_{\ds}}$
decreases. In addition, although $g_{\ds}$ does not depend on $P_{\Delta\Delta}$, the $g_{_{\ds\Delta\Delta}}$ is proportional to the square
root of the channel probability $\sqrt{P_{\Delta\Delta}}$. Finally, we would mention that we cannot dynamically determined the size
parameter as well as the probability in this approach. Instead, to proceed the calculation without contradicting the results given by
the quark model, we simply borrow the corresponding qualitative conclusions given in those dynamical quark model calculations.\\

\subsection{Cross section for $p\bar{p}\to\Delta\bar{\Delta}\to \ds\bar{\ds}$}

In the CM energy region of $\sqrt{s}\in[4.8-5.5]~GeV$, the evaluated total cross section of the process $p\bar{p}\to\Delta\bar{\Delta}\to
\ds\bar{\ds}$, shown by the Feynman diagram in Fig. 3,  is given in  Fig. 5. Here, we reiterate that the cross section is evaluated based
on the qualitative interpretations of $\ds$ in the non-relativistic  quark model approach, with which all observed
properties of $\ds$ can be well described. The cross section curve in Fig. 5 tells us that the total cross section in the
$d^*\bar{d^*}$ pair production process is about 4-6 orders of magnitude smaller than that in the $p\bar{p}\to\Delta\bar{\Delta}$ reaction.\\

\vspace{0.5cm}
\begin{figure}[h]
\centering
\includegraphics [width=8cm, height=6cm]{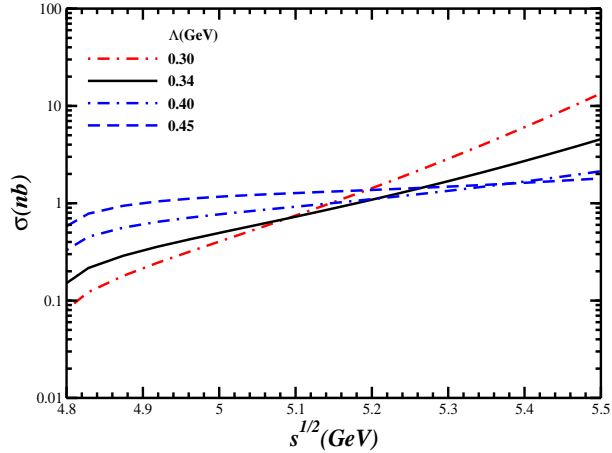}
\caption{The estimated cross section for the reaction of
$p\bar{p}\to \ds\bds$ in units of $(nb)$. }
\end{figure}

We know that the cross section in Fig.5 is dependent on the phase space as well as the matrix element ${\cal M}_{if}$. The phase space
increases with the increasing $\sqrt{s}$. The matrix element of ${\cal M}_{if}$ relates to model parameter $\Lambda$ as well as to
$\sqrt{s}$. The estimated total cross section of $p\bar{p}\to\Delta\bar{\Delta}\to \ds\bar{\ds}$ is also subject to
the impact of the interpretation of the $\ds$ state,  namely its size and the probability of its $\Delta\Delta$ component. The
resultant cross section (with a fixed value of $P_{\Delta\Delta}\sim 1/3$) in Fig. 5 shows its dependence on $\Lambda$. Actually the
coupling $g_{_{\ds\D\D}}$ is proportional to $\sqrt{P_{\Delta\Delta}}$ and the matrix element ${\cal M}_{if}$ is
proportional to $P_{\Delta\Delta}$. Thus, the obtained cross section changes with respect to $P^{2}_{\Delta\Delta}$. Moreover, the
coupling $g_{_{\ds\D\D}}$ and the matrix element ${\cal M}_{if}$ are closely related to the structure of the mass operator in the
structural integral and to the loop calculations of Fig. 3, respectively. Here, we only display the $\Lambda$ dependence
explicitly in Fig. 5. It shows that in the small $\sqrt{s}$ region, say less than $5.2~GeV$, the cross section is distinctly suppressed,
because the coupling constant $g_{d^*}$ decreases due to the increase of $\Lambda$. However, when $\sqrt{s}$ is greater than
$5.5~GeV$, the production cross section with a smaller $\Lambda$ value, say less than $0.34~GeV$, may increase dramatically
with the increase of $\sqrt{s}$ due to the larger structural integral, caused by a larger $\Lambda$-dependent $g_{_{\ds\D\D}}$
value, and a larger phase space. It should be reiterated that as a rough estimate, we only consider the production
cross section for the $\ds$-$\bar{\ds}$ pair in this paper, and not take the complicated background contribution into account. When the
CM energy is about $5.2~GeV$, the $\Lambda$ dependence of the cross section becomes small, and the estimated cross section becomes meaningful.\\

It should be noted that the obtained cross section of $p\bar{p}\to\Delta\bar{\Delta}\to \ds\bar{\ds}$ is in the order of
$nb$. According to the designed luminosity and integrated luminosity of $\dP$Danda, which are about $\sim 2\times 10^{32}cm^{-2}/s$ and
$\sim 10^{4}nb^{-1}/day$, respectively, we expect that about $(0.51, 0.71, 1.19)\times 10^{4}$ $\ds\bar{\ds}$ events can be
observed per-day at $\sqrt{s}=(5.0, 5.1, 5.2)~GeV$,  if the overall efficiency is $100\%$. On the other hand, from a
technical point of view, $\ds$ cannot be directly observed. Observation of $\ds$ is usually achieved through the measurements of its strong
decay processes, namely measuring various mesons and baryons, such as $\pi$, proton, neutron and etc., and measuring some invariant mass
spectra and Dalitz plots  and etc.. It is noticed that the dominated decay channels of $\ds$ are $\ds\to d\pi\pi$ and $\ds\to pn\pi\pi$ with
their partial decay widths of about $27~MeV$ and $31~MeV$, respectively, which correspond to the branching ratios of about $36\%$ and
$41\%$, respectively. As a consequence, the possible events of $p\bar{p}\to\ds\bar{\ds}\to \bar{\ds}\,~d\pi\pi$
(or $p\bar{p}\to\ds\bar{\ds}\to \ds\,\bar{d}\pi\pi$) and $p\bar{p}\to\ds\bar{\ds}\to \bar{\ds}\, pn\pi\pi$ (or $p\bar{p}\to\ds\bar{\ds}\to \ds\,
\bar{p}\bar{n}\pi\pi$) can roughly be estimated. They are respectively about $((0.18,0.21), (0.26,0.29), (0.43,0.49))\times 10^{4}$ per-day at
$\sqrt{s}=(5.0,5.1,5.2)~GeV$ (if the overall efficiency is being assumed 100\%). Finally, it should be further mentioned that in order to
avoid the interference caused by the background of a large number of produced pions and nucleons, according to our previous
discussion~\cite{Lu:2018gtk,Lu:2018wsr}, it may be more practical to confirm the existence of $\ds$ by looking for $\bar{\ds}$ via
the decay channels in above brackets.\\

\section{Summary and discussions}

In this work, we estimate the cross section of the $p\bar{p}\to\Delta\bar{\Delta}\to \ds\bar{\ds}$ reaction, which
might possibly be measured at forthcoming experiments at $\bar{P}$anda in the CM energy of $\sqrt{s}\in [4.8, 5.5]~GeV$. A
relativistic and covariant phenomenological effective Lagrangian approach is employed in the practical calculation.  To describe the
structure of the outgoing $\ds\bds$ pair, qualitative conclusions from the sophisticated and dynamical calculations in the
non-relativistic constituent quark model, with which all existing data can be well explained, are directly adopted to
fix the model-parameter $\Lambda$ approximately. The estimated production cross section for $\ds\bds$ should be a lower
bound, since in our assumption, only 1/3 of $\Delta\Delta$ is considered to be an ingredient of $\ds$. The result shows that the estimated
production cross section of this reaction is in the order of $nb$ which is much smaller than the known cross section of
$p\bar{p}\to\Delta\bar{\Delta}$ whose value is  in the order of $mb$. Nevertheless, among a huge amount of events of produced
hadron pair at $\bar{P}$anda, there may still exist a certain amount of events of produced $\ds\bds$ pair. These events are expected to
be observed through measuring the final baryons and mesons in some strong decay processes of $\ds$, such as $\ds\to d\pi\pi$
(or $\bar{\ds}\to \bar{d}\pi\pi$) and $\ds\to pn\pi\pi$ (or $\bar{\ds}\to \bar{p}\bar{n}\pi\pi$). We also rough estimate the
possible events of these processes from the branching ratios of the $\ds$ strong decays as a reference.\\

\end{CJK*}

\newpage
\begin{thebibliography}{99}

\bibitem{Clement:2016vnl}
  H.~Clement, "On the History of Dibaryons and their Final Observation,", arXiv:1610.05591 [nucl-ex],
  Prog.\ Part.\ Nucl.\ Phys.\  {\bf 93}, 195 (2017).

\bibitem{Clement:2020mab} H. Clement and T. Skorodko, "Dibaryons: Molecular versus Compact Hexaquarks",
arXiv:2008.07200 [nucl-th], Chin. Phys. C {\bf 45}, 022001 (2021).

\bibitem{CELSIUS-WASA}M. Bashkanov et al., Phys. Rev. Lett. {\bf 102} 052301 (2009).

\bibitem{Adlarson:2011bh}
  P.~Adlarson {\it et al.} [WASA-at-COSY Collaboration],
"ABC Effect in Basic Double-Pionic Fusion --- Observation of a new resonance,",
  Phys.\ Rev.\ Lett.\  {\bf 106}, 242302 (2011).

\bibitem{Adlarson:2012fe}
  P.~Adlarson {\it et al.} [WASA-at-COSY Collaboration],
"Isospin Decomposition of the Basic Double-Pionic Fusion in the Region of the ABC Effect,"
  Phys.\ Lett.\ B{\bf 721}, 229 (2013).

\bibitem{Adlarson:2014pxj}
  P.~Adlarson {\it et al.} [WASA-at-COSY Collaboration],
  Phys.\ Rev.\ Lett.\  {\bf 112}, no. 20, 202301 (2014).

\bibitem{Chen:2016qju}
  H.~X.~Chen, W.~Chen, X.~Liu and S.~L.~Zhu, "The hidden-charm pentaquark and tetraquark states," arXiv:1601.02092 [hep-ph],
  Phys.\ Rept.\  {\bf 639}, 1 (2016).

\bibitem{Guo:2017jvc}
  F.~K.~Guo, C.~Hanhart, U.~G.~Meissner, Q.~Wang, Q.~Zhao and B.~S.~Zou, "Hadronic molecules,"
  arXiv:1705.00141 [hep-ph], Rev.\ Mod.\ Phys.\  {\bf 90}, 015004 (2018).

\bibitem{Dong:2017gaw} Y.~Dong, A.~Faessler and V.~E.~Lyubovitskij,
  Prog.\ Part.\ Nucl.\ Phys.\  {\bf 94} (2017) 282.

\bibitem{Lebed:2016hpi}
  R.~F.~Lebed, R.~E.~Mitchell and E.~S.~Swanson, "Heavy-Quark QCD Exotica," arXiv: 1610.04528 [hep-ph],
  Prog.\ Part.\ Nucl.\ Phys.\  {\bf 93}, 143 (2017).

\bibitem{Yuan} X. Q. Yuan et al., Phys. Rev. C {\bf 60}, 045203 (1999).

\bibitem{Brodsky}M. Bashkanov, Stanley J. Brodsky, and H. Clement,
Phys. Lett. B {\bf 727}, 438 (2013).

\bibitem{Huang:2014kja}
  F.~Huang, Z.~Y.~Zhang, P.~N.~Shen and W.~L.~Wang,
  Chin.\ Phys.\ C {\bf 39}, no. 7, 071001 (2015).

\bibitem{Huang:2015nja}
  F.~Huang, P.~N.~Shen, Y.~B.~Dong and Z.~Y.~Zhang,
  Sci.\ China Phys.\ Mech.\ Astron.\  {\bf 59}, no. 2, 622002 (2016).

\bibitem{Dong}Yubing Dong, Pengnian Shen, Fei Huang, and Zongye Zhang,
Phys. Rev. C{\bf 91}, 064002 (2015).

\bibitem{Dong1}Yubing Dong, Fei Huang, Pengnian Shen, and Zongye Zhang,
Phys. Rev. C{\bf 94}, 014003 (2016).

\bibitem{Dong2}Yubing Dong, Fei Huang, Pengian Shen, and Zongye Zhang, Chinese Physics C{\bf 41} (2017) 101001.

\bibitem{Huang:2013nba}
  H.~Huang, J.~Ping and F.~Wang, "Dynamical calculation of the $\Delta\Delta$ dibaryon candidates," arXiv:1312.7756 [hep-ph],
  Phys.\ Rev.\ C {\bf 89}, no. 3, 034001 (2014).

\bibitem{Huang:2018rpb}F. Huang, and W. L. Wang, "Nucleon-nucleon interaction in a chiral SU(3) quark model revisited",
Phys. Rev. D, {\bf 98}, No. 7, 074018, 2018.


\bibitem{Gal:2013dca}  A.~Gal and H.~Garcilazo,
  Phys.\ Rev.\ Lett.\  {\bf 111}, 172301 (2013).

\bibitem{Gal:2014zia}
  A.~Gal and H.~Garcilazo, Nucl.\ Phys.\ A {\bf 928}, 73 (2014).

\bibitem{Platonova:2014rza}
  M.~N.~Platonova and V.~I.~Kukulin, "Hidden dibaryons in one- and two-pion production in NN collisions," arXiv:1412.4574 [nucl-th]
  Nucl.\ Phys.\ A {\bf 946}, 117 (2016).

\bibitem{Shi:2019dpo}Pan-Pan Shi, Fei Huang, and Wen-Ling Wang, "$d^*(2380)$ and its partners in a diquark model",
Eur. Phys. J. C, {\bf 79}, No. 4, 314 (2019).

\bibitem{Ikeno:2021frl} Natsumi Ikeno, Raquel Molina, and Eulogio Oset,  "Triangle singularity mechanism for the $pp\rightarrow\ensuremath{\pi}+d$
fusion reaction",
Phys. Rev. C, {\bf 104}, No. 1, 014614, (2021).

\bibitem{Platonova:2012am}
  M.~N.~Platonova and V.~I.~Kukulin, "ABC-effect as a signal of chiral symmetry restoration in hadronic collisions,"
  arXiv: 1211.0444 [nucl-th],  Phys.\ Rev.\ C {\bf 87}, no. 2, 025202 (2013).

\bibitem{Dong:2017mio}
Yubing Dong, Fei Huang, Pennian Shen and Zongye Zhang,
Phys.\ Rev.\ D {\bf 96}, 094001 (2017).

\bibitem{Dong:2018emq}
Yubing Dong,  Pennian Shen and Zongye Zhang,
Phys.\ Rev.\ D {\bf97}, no. 11, 114002 (2018).

\bibitem{Dong:2019gpi}
Yubing Dong, Fei Huang, Pennian Shen and Zongye Zhang,
Int.\ J.\ Mod.\ Phys.\  A {\bf 34}, no. 18, 1950100 (2019).

\bibitem{Ishikawa:2016yiq}
T. Ishikawa  {\it et al.},
"First measurement of coherent double neutral-pion photoproduction on the deuteron at incident energies below 0.9 GeV",
arXiv: 1610.05532 [nucl-ex], Phys.\ Lett.\ B {\bf 772}, 398--402 (2017).

\bibitem{Ishikawa:2018wkv}
T. Ishikawa  {\it et al.}, "Non-strange dibaryons studied in the $\gamma d\to \pi^0\pi^0 d$ reaction",
Phys.\ Lett.\ B {\bf 789}, (2019), 413--318.

\bibitem{Ishikawa:2020jsh}
T. Ishikawa  {\it et al.}, "Study of Non-strange Dibaryon Resonances Via Coherent Double Neutral-Meson Photoproduction from the
Deuteron," Springer Proc. Phys., 238 (2020), 609-613.

\bibitem{Fioravanti:2014vza} Elisa Fioravanti,
"Experimental overview of the $\overline{P}$anda experiment", J. Phys. Conf. Ser. {\bf 503}, 012030 (2014).

\bibitem{Hawranek:2007mp}
P. Hawranek, "Hadron physics experiments in antiproton proton reactions with the planned PANDA detector", Int. J. Mod. Phys. A {\bf 22},
574-577 (2007).

\bibitem{Bettoni:2012zz} Diego Bettoni,  "Hadron physics with the PANDA experiment at the FAIR", Prof. Part. Nucl. Phys.
{\bf 67}, 502-510 (2012).

\bibitem{Dong:2008gb}
  Y.~B.~Dong, A.~Faessler, T.~Gutsche and V.~E.~Lyubovitskij,
"Estimate for the X(3872) to gamma $J/\psi$ decay width,"
  Phys.\ Rev.\  D {\bf 77}, 094013 (2008);

\bibitem{Dong:2009yp}
  Y.~B.~Dong, A.~Faessler, T.~Gutsche, S.~Kovalenko and V.~E.~Lyubovitskij,
"X(3872) as a hadronic molecule and its decays to charmonium states and pions,"
  Phys.\ Rev.\  D {\bf 79}, 094013 (2009).
\bibitem{Dong:2009uf}
Yubing~Dong, Amand~Faessler, Thomas~Gutsche, and Valery~E.~Lyubovitskij,
"$J/\psi$ gamma and $\psi(2S)$ gamma decay modes of the X(3872)", arXiv:0909.0380 [hep-ph], J. Phys. G. {\bf 38}, (2011), 015001.

\bibitem{Dong:2013iqa}
Yubing~Dong, Amand~Faessler, Thomas~Gutsche, and Valery~E.~Lyubovitskij, "Strong decays of molecular states Z$_{c}^{+}$ and
Z$_{c}^{'+}$", arXiv:1306.0824 [hep-ph], Phys. Rev. D {\bf 88}, (2013), 014030.

\bibitem{Dong:2009tg}
  Y.~Dong, A.~Faessler, T.~Gutsche and V.~E.~Lyubovitskij,
"Strong two-body decays of the $\Lambda_c(2940)+$ in a hadronic molecule picture," arXiv:0910.1204 [hep-ph]
  Phys.\ Rev.\ D {\bf 81}, 014006 (2010).

\bibitem{Dong:2010xv}
Y.~Dong, A.~Faessler, T.~Gutsche, S.~Kumano and V.~E.~Lyubovitskij,
"Radiative decay of $\Lambda_c(2940)^+$ in a hadronic molecule picture"
  Phys.\ Rev.\ D {\bf 82}, 034035 (2010).

\bibitem{Dong:2008mt}
Y.~Dong, A.~Faessler, T.~Gutsche, S.~Kumano and V.~E.~Lyubovitskij,
"Phenomenological Lagrangian approach to the electromagnetic deuteron form factors",
arXiv:0806.3679 [hep-ph], Phys. Rev. C {\bf 78}, 035205 (2008).

\bibitem{Faessler:2003yf} Amand Faessler, Thomas Gutsche, M. A. Ivanov, Valery E. Lyubovitskij and P. Wang,
"Pion and sigma meson properties in a relativistic quark model", arXiv:0304031[hep-ph], Phys. Rev. D {\bf 68}, 014011(2003).

\bibitem{Xiao:2020alj}
Cheng-Jian Xiao, Yubing Dong, Thomas Gutsche, Valery E.
Lyubovitskij, and Dian-Yong Chen, "Towards the decay properties of
deuteron-like state $d_{N\Omega}$", arXiv: 2004.12415 [hep-ph],
Phys.\ Rev. D {\bf 101}, (2020), 114032.


\bibitem{Lis:2000}
Q. B. Li, P.N.Shen, "$N\Omega$ and $\Delta\Omega$ dibaryons in a SU(3)chiral quark model", Euro.\ Phys.\ J.\ A{\bf 8},(2000), 417.


\bibitem{Iritani:2018sra} T.~Iritani {\it et al.} (HAL QCD Collaboration),
"$N\Omega$ dibaryon from lattice QCD near the physical point",
Phys.\ Lett.\ B {\bf 792}, 284 (2019).

\bibitem{Scadron:1968zz}
D. Michael Scadron, "Covariant Propagators and Vertex Functions for Any Spin", Phys.\ Rev.\ {\bf 165}, (1968), 1640-1647.

\bibitem{Salam:1962ap} A.~Salam, "Lagrangian Theory Of Composite Particles,"
  Nuovo Cim.\  {\bf 25}, 224 (1962).

\bibitem{Weinberg:1962hj}S.~Weinberg, "Elementary Particle Theory Of Composite Particles,"
  Phys.\ Rev.\  {\bf 130}, 776 (1963).

\bibitem{Hayashi:1967hk}K.~Hayashi, M.~Hirayama, T.~Muta, N.~Seto and T.~Shirafuji,
  Fortsch.\ Phys.\ {\bf 15}, 625 (1967).

\bibitem{Efimov:1993ei} G.~V.~Efimov and M.~A.~Ivanov,
  {\it The Quark Confinement Model of Hadrons},
  (IOP Publishing, Bristol $\&$ Philadelphia, 1993).

\bibitem{Hayashi:1967bjx} Kenji Hayashi,  Minoru Hirayama, Taizo  Muta,  Noriaki Seto, Noriaki, and  Takeshi Shirafuji,
"Compositeness criteria of particles in quantum field theory and S-matrix theory",
Fortsch. Phys., {\bf 15}, No. 10, 625-660, 1967.

\bibitem{Efimov:1987sa}G. V. Efimov, M. A. Ivanov, and V. E. Lyubovitskij,
"Strong nucleon and $\Delta$-isobar form factors in quark confinement model",
Few Body Syst., {\bf 6}, 17-43, 1989.

\bibitem{VanApeldoorn:1978jd}
G. W. Van Apeldoorn, R. L. F. Gruendeman, D. Harting, {\it et al}., "Final States with a Neutral Meson and a $\pi^+\pi^-$ or
$\bar{p}p$ Pair in anti-p p Interactions at 7.2-GeV/c", Nucl. Phys. B. {\bf 133} 245--265 (1978).

\bibitem{Ward:1978npb}D. R. Ward?? R. E. Ansorge?? C. P. Bust, et al., "$\D^{++}$ and $\overline{\D^{++}}$ production in 100 GeV/c
$\bar{p}p$ interactions", Nucl. Phys. B. {\bf 141}, 203-219 (2978).

\bibitem{Johnson:1980nw}P. Johnson, et al., "General Features of the Anti-p $p$ Interaction at 12-{GeV}/c",  Nucl. Phys. B {\bf 173}, 77--92 (1980).

\bibitem{VanApeldoorn:1982iw} G. W. Van Apeldoorn  {\it et al.}, Z.\ Phys.\ C {\bf 12}, (1985), 95--98.

\bibitem{Saleem:1983ky}Mohammad Saleem and Fazal-e-Aleem, Prog.\ Theor.\ Phys. {\bf 70}, 1156, (1983).

\bibitem{Wolf:1969iw} E. Guenter Wolf, Phys.\ Rev.\ {\bf 182}, (1969), 1538--1560.

\bibitem{Cao:2010ji}
Xu Cao, Bing-Song Zou, Bing-Song, Hu-Shan Xu,  "Phenomenological study on the $\bar p N\to \bar NN\pi\pi$ reactions,"
arXiv: 1009.1060 [nucl-th], Nucl.\ Phys.\ A {\bf 861}, (2011), 23--36.

\bibitem{Riska:2000gd}
D. O. Riska and G. E. Brown, "Nucleon resonance transition couplings to vector mesons,"
arXiv:0005049 [nucl-th], Nucl.\ Phys.\ A {\bf 679}, (2011), 577-596.

\bibitem{Lu:2018gtk} Chao-Yi L\"u, Ping Wang, Yubing Dong, Peng-nian Shen, and Zong-ye Zhang,
"The opportunity to find $\bar{d}^\ast(2380)$ in the $\Upsilon(nS)$ decay", arXiv: 1810.02138 [hep-ph],
Phys. Rev. D {\bf 99}, 036015 (2019).

\bibitem{Lu:2018wsr} Chao-Yi L\"u, Ping Wang, Yubing Dong, Peng-nian Shen, Zong-ye Zhang, and D. M. Li,
"Phenomenological study on the decay widths of $\Upsilon(nS)\rightarrow \bar{d}^{\ast}(2380)+X$", arXiv: 1803.07795 [hep-ph],
Chinese Phys. C {\bf 42}, 064102 (2018).
\end{thebibliography}
\end{document}